\documentclass[12pt,a4paper]{article}

\usepackage{amsmath,amssymb,epsfig}

\begin{document}

\newcommand{\el}{\left}
\newcommand{\er}{\right}
\newcommand{\p}{\prime}
\newcommand{\ka}{\kappa}
\newcommand{\bc}{\beta C}
\newcommand{\rr}{\rho}
\newcommand{\ti}{\tilde}
\newcommand{\veps}{\varepsilon}
\newcommand{\dis}{\displaystyle}
\newcommand{\scr}{\scriptsize}

\title{MICROSCOPIC MODEL ANALYSIS \\
OF THE $^{6}$He,~$^{6}$Li\,+\,$^{28}$Si TOTAL REACTION CROSS
SECTIONS AT THE ENERGY RANGE 5-50\,A MEV}

\author{K. V. LUKYANOV, I. N. KUKHTINA, V. K. LUKYANOV, \\ Yu. E. PENIONZHKEVICH,
Yu. G. SOBOLEV, E. V. ZEMLYANAYA}

\date{}
\maketitle

\begin{center}
Joint Institute for Nuclear Research, Dubna 141980, Russia
\end{center}

\begin{abstract}
The existing and some preliminary experimental data on the total
cross sections of the $^{4,6}$He,~$^{6,7}$Li\,+$^{28}$Si reactions
at energies E=5-50 A MeV are demonstrated. The data on
$^{6}$Li,$^{6}$He+$^{28}$Si are analyzed in the framework of the
microscopic optical potential with real and imaginary parts
obtained with a help of the double-folding procedure and by using
the current models of densities of the projectile nuclei .
Besides, the microscopic double-folding Coulomb potential is
calculated and its effect on cross sections is compared with that
when one applies the traditional Coulomb potential of the uniform
charge distribution. The semi-microscopic potentials are
constructed from both the renormalized microscopic potentials and
their derivatives to take into account collective motion effect
and to improve an agreement with experimental data.
\end{abstract}

\section{Introduction}

Generally, the aim of our study is to analyze the possibility of
the microscopic optical potential to give a physical
interpretation of the total reaction cross sections of
$^{4,6}$He,~$^{6,7}$Li on $^{28}$Si (see refs.
\cite{Kuz}-\cite{Sob}) including some preliminary data on the
$^{6}$Li+$^{28}$Si reaction at the energies E=5-50 A MeV (see
Fig.~1), and at present we study only the $^{6}$He,$^{6}$Li
$+^{28}$Si cross section. There is the following motivation of
this task. First, an interpretation of experimental data with a
help of usually applied phenomenological optical potentials does
not answer questions both on the nuclear structure of colliding
nuclei and on the mechanism of their scattering. Moreover, such
kind of fitting is, in fact, only the parametrization of data by
introducing a set, say, of the six or more free parameters, which
are different for different energies and kinds of interacting
nuclei.

\begin{figure}[t]
\begin{center}
\psfig{file=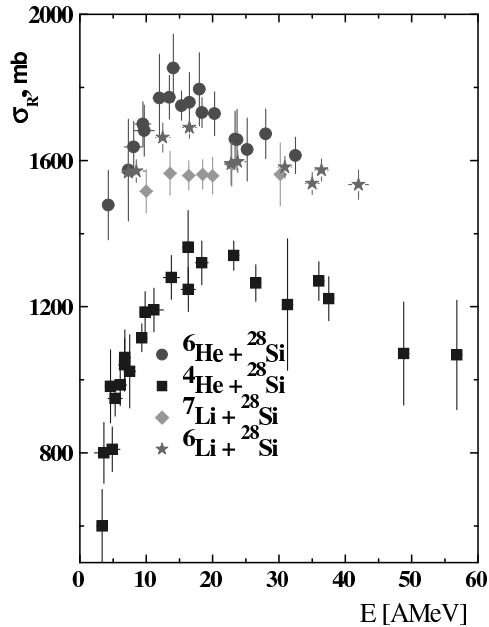,width=2.5in}
\end{center}
\caption{The total reaction cross sections measured in
\cite{Kuz}-\cite{Sob}.} \label{fig1}
\end{figure}

Otherwise, the microscopic models do not contain free parameters
and provide the possibility to test the models of nuclear
structure. Particularly, in this paper we use the current models
of the projectile nuclei $^6$He and $^6$Li to estimate a
sensitivity of total cross sections to a behavior of their
densities in the peripheral region. In calculations, we use the
Tanihata model \cite{Tan} and the cluster-orbital shell-model
approximation (COSMA) \cite{Zhuk} of density distributions of bare
protons (Z) and neutrons (N) in nuclei
%(1.1)
\begin{equation}\label{eq1.1}
\rr_X(r)={2\over (\bar a\sqrt{\pi})^3}\,e^{\dis{-(r/\bar a)^2}} +
{X-2\over 3}{2\over (\bar b\sqrt{\pi})^3}\el({r\over\bar b}\er)^2\,
e^{\dis{-(r/\bar b)^2}}, \qquad  X=Z,N,
\end{equation}
where $\bar a,~\bar b$ are parameters of the models. Also,
densities of the large-scale shell-model (LSSM) \cite{Karat} is
also presented together with the $^6$Li density from Tables of \ \
\cite{Patt} . In Fig.~2 one sees the visual distinction of shapes
of the proton, neutron and the nuclear matter densities obtained
in these models. Between these densities only the LSSM has the
realistic exponential  behavior at large distances while the
others have the Gaussian shape of tails.

\begin{figure}[t]
\begin{center}
\psfig{file=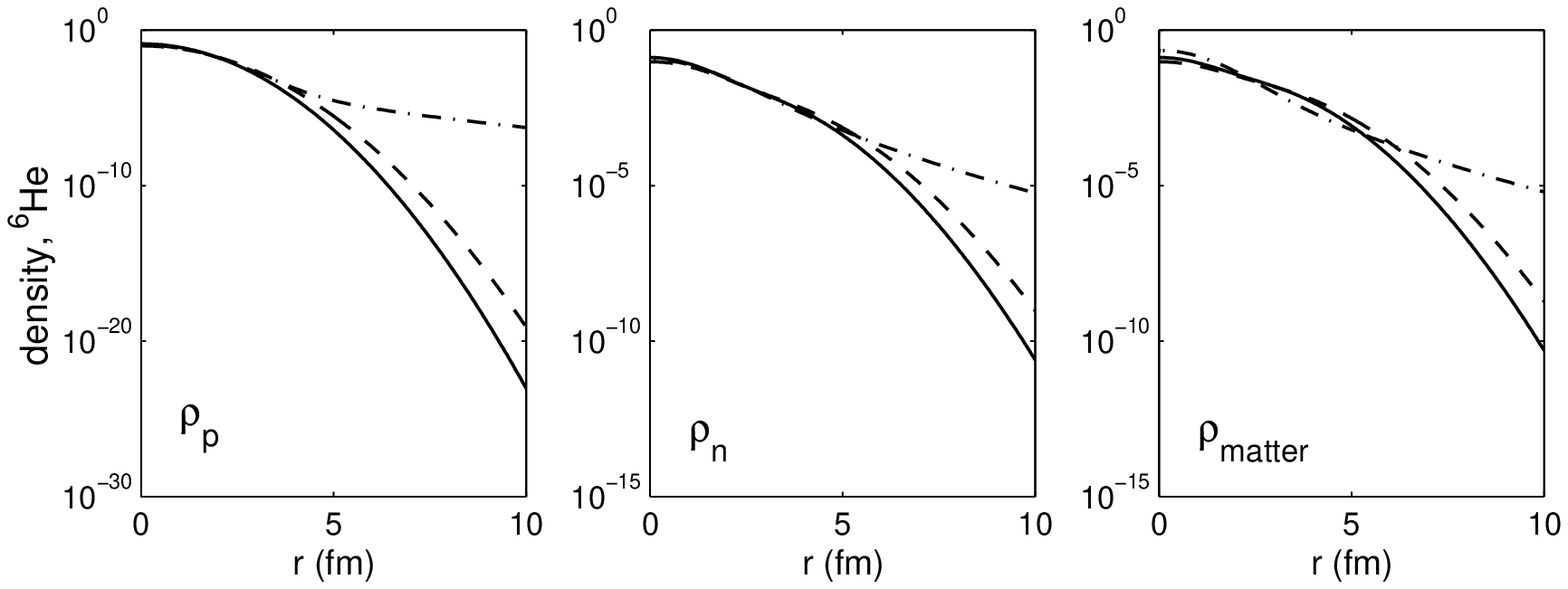,width=.9\linewidth}
\psfig{file=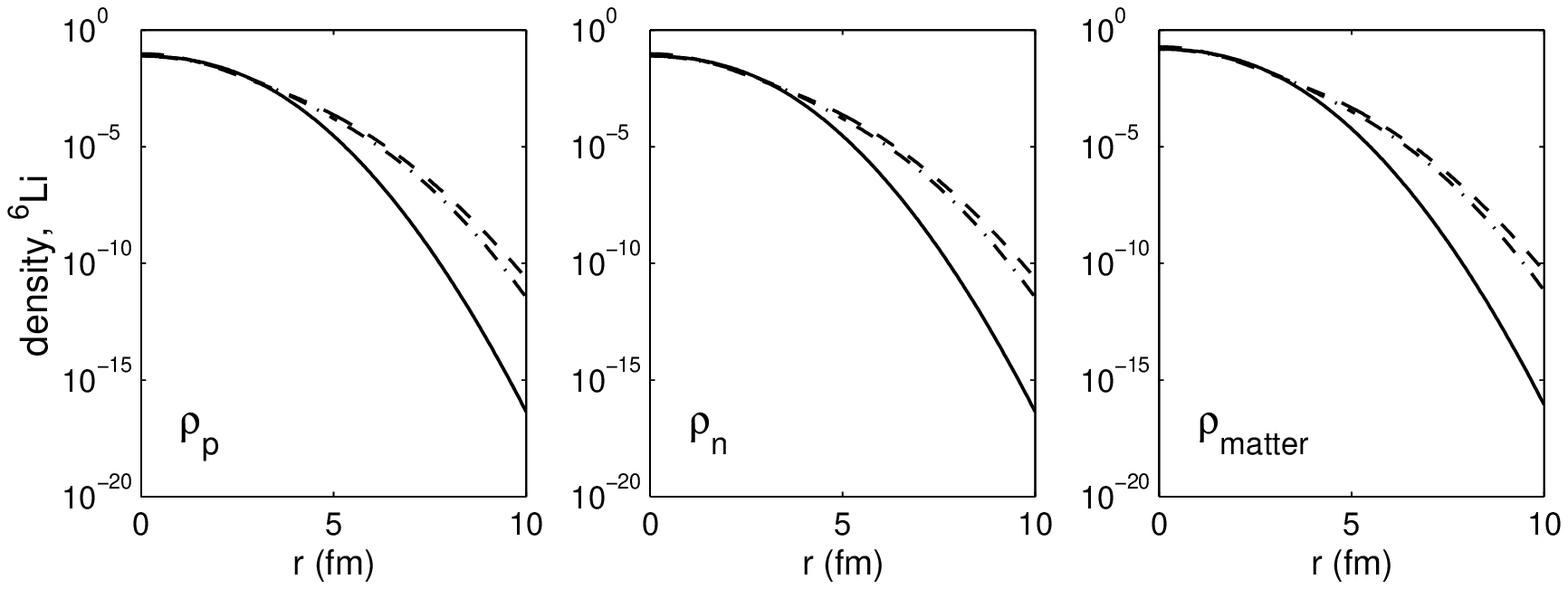,width=.9\linewidth}
\end{center}
\caption{Density distributions of $^{6}$He  and $^{6}$Li,
calculated in
 different models. Solid curves: Tanihata ($^6$He) and table ($^6$Li); dash-dotted: LSSM; dashed: COSMA (see the text)  }
\label{fig2}
\end{figure}

In Sec.~2, the double-folding model (see, e.g.,\cite{Khoa}),
including the exchange term, is applied to calculate the real part
of the microscopic optical potential whereas for its imaginary
part we takes the form obtained in \cite{LZL} basing on
high-energy approximation (HEA) theory of scattering
\cite{{G},{S}} . Applications of the microscopic potentials are
made in Sec.~3. The role of the Coulomb potential is also analyzed
by comparison of  cross sections calculated with  the traditional
Coulomb potential of the uniform charge density distribution, and
with that obtained in the framework of folding procedure
accounting for the realistic  nuclear charge density
distributions. Then, we discuss the method of adding free
parameters to account for influence of collective modes of nuclei.
Summary and conclusion are done in Sec.~4.

\section{Microscopic optical potential}

The double-folding nucleus-nucleus potential (the real one)
consists of the direct and exchange parts:
%[2.1]
\begin{equation}\label{eq2.1}
V^{DF}=V^D~+~V^{EX}
\end{equation}
%[2.2]
\begin{equation}\label{eq2.2}
\qquad V^D(r) = \int d^3 r_p d^3 r_t \, \rr_p({\bf r}_p)\,
\rr_t({\bf r}_t)\, v_{NN}^D(s), \qquad {\bf s}={\bf r}+{\bf
r}_t-{\bf r}_p,
\end{equation}
%[2.3]
$$V^{EX}(r) = \int d^3 r_p \, d^3 r_t \, \rr_p({\bf r}_p, \,{\bf r}_p+
{\bf s}) \, \rr_t({\bf r}_t, {\bf r}_t-{\bf s})\times$$
\begin{equation}\label{eq2.3}
\times v_{NN}^{EX}(s)\, \exp\el[{i{\bf K}(r)\cdot s\over M}\er],
\end{equation}
where $\rr_{p,t}$ are the one-particle projectile (p) and  target
(t) matrices of densities. The modern calculations are usually
apply the effective Paris nucleon-nucleon CDM3Y6 potential
$v_{NN}$ having the form
%[2.4]
\begin{equation}\label{eq2.4}
v_{NN}(E,\rr, s)=g(E)\,F(\rr)\,v(s), \quad
v(s)=\sum\limits_{i=1,2,3}N_i{\exp(-\mu_i s)\over \mu_i s},
\end{equation}
where the energy and density dependencies are given as
%[2.5]
\begin{equation}\label{eq2.5}
g(E)=1-0.003E/A_p,\,\,\, F(\rr)=C\Bigl
[1+\alpha\exp(-\beta\rr)-\gamma\rr\Bigr],\,\,\, \rr=\rr_p+\rr_t,
\end{equation}
$$
C=0.2658,\quad \alpha=3.8033, \quad \gamma=4.0,
$$
and the parameters $N_i$ and $\mu_i$ are done in \cite{Khoa} . The
energy dependence of $V^{EX}$ arises primarily from the
contribution the exponential in the integrand, where
$K(r)=\{2Mm/\hbar^2[E-V_N^{DF}(r)- V_c(r)]\}^{1/2}$ is the local
nucleus-nucleus momentum, $M=A_pA_t/(A_p+A_t)$, $m$ is the nucleon
mass, and therefore there occurs the typical non-linear problem.

Here we paid an attention on the important role of the exchange
effect in calculations of nucleus-nucleus real potentials. This is
depicted in Fig.~3 where the double-folding $V^{DF}$-potential for
the $^6$He+$^{28}$Si scattering at E=25 MeV/nucleon is calculated
with a help of two different every so often kinds of effective
$v_{NN}$ potentials, the Paris CDM3Y6 and the Reid DDM3Y1
potentials. They have different sets of the parameters
$N_i,\,\mu_i$ and $C,\,\alpha,\,\beta,\,\gamma$ (see \cite{Khoa}).
It is seen, that their direct parts has different signs, and thus
the exchange part plays the crucial role in forming the whole
nuclear potential.

\begin{figure}
\begin{center}
\psfig{file=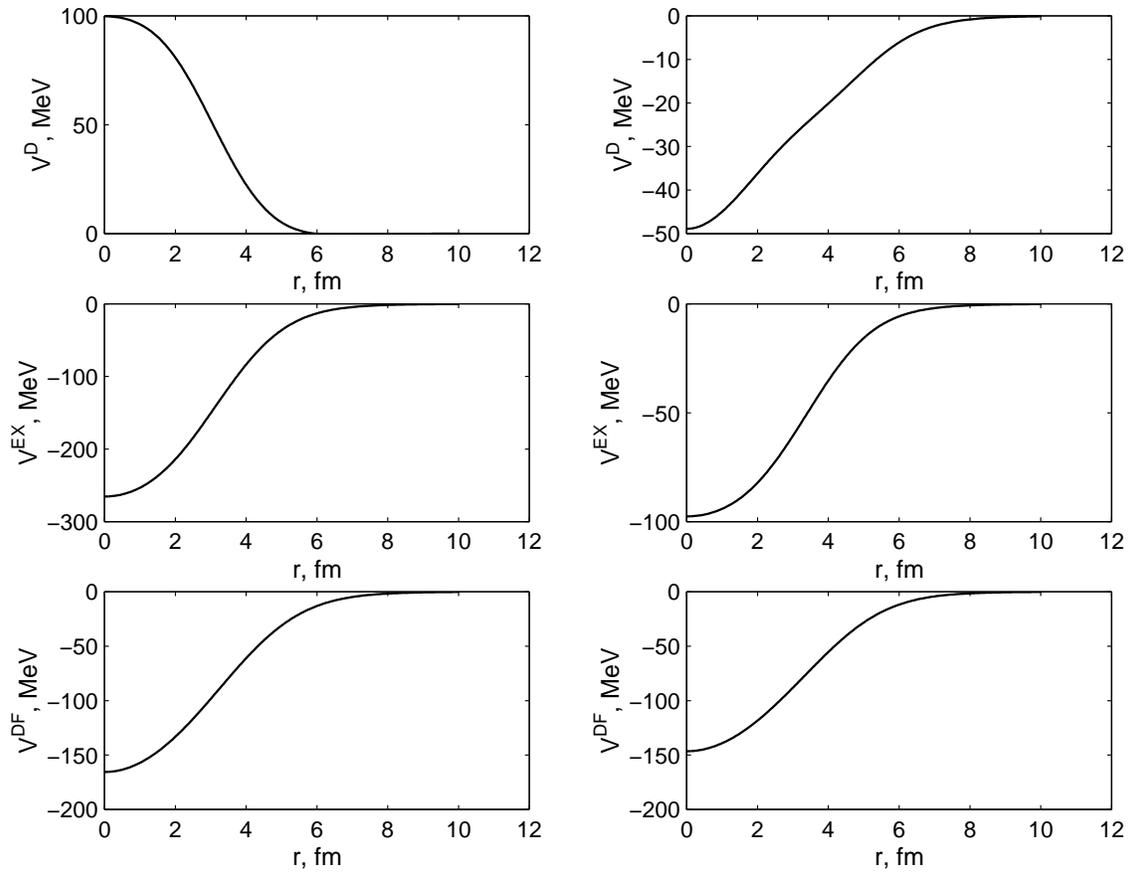,width=.9\linewidth}
\end{center}
\caption{Behavior of different terms and of the total
nucleus-nucleus potentials of $^6$He$+^{28}$Si calculated with the
Paris (left side) and Reid (right side) effective NN-potentials
(see the text).} \label{fig3}
\end{figure}

Note that when constructing microscopic optical potentials people
usually use only the real double-folding potential
(\ref{eq2.1})-(\ref{eq2.4}) while the imaginary part is taken in a
 phenomenological form with free parameters fitted to experimental
data for each specific energy individually. Instead, in our
calculations we use below the imaginary part as it is done in the
microscopic optical potential (HEA-potential) obtained in
\cite{LZL} basing on the HEA theory  \cite{{G},{S}} . Its
imaginary part is as follows:
%[2.6]
\begin{equation}\label{eq2.6}
W^H(r)\,=\,-{2E\over k(2\pi)^2}{\bar\sigma}_{NN}
\int_0^\infty dq~q^2j_0(qr){\ti\rr}_p(q){\ti\rr}_t(q){\ti f}_N(q),
\end{equation}
where ${\ti\rr}(q)=\int d^3r~\exp(i{\bf q}{\bf r}) \rr(r)$  is the
form factor of a pointlike nuclear density, and $\sigma_{NN}$ is
the total nucleon-nucleon cross sections that is  parametrized in
\cite{CG} as a function of the NN collision energy. The
superscript H indicates the HEA roots of the potential.

\section{Results of the cross section calculations}

Fig.~4 exhibits microscopic calculations of the total reaction
cross sections based on the given folding potentials
(\ref{eq2.2},\ref{eq2.3},\ref{eq2.6}) and using density
distributions of the projectile nuclei $^6$He and $^6$Li. It is
seen that they exceed the experimental data and their shapes
follow to the data at energies higher than 15 Mev/nucleon.
Numerical calculations of cross sections were made by using the
code DWUCK4 \cite{Kunz} .

\begin{figure}[t]
\begin{center}
\psfig{file=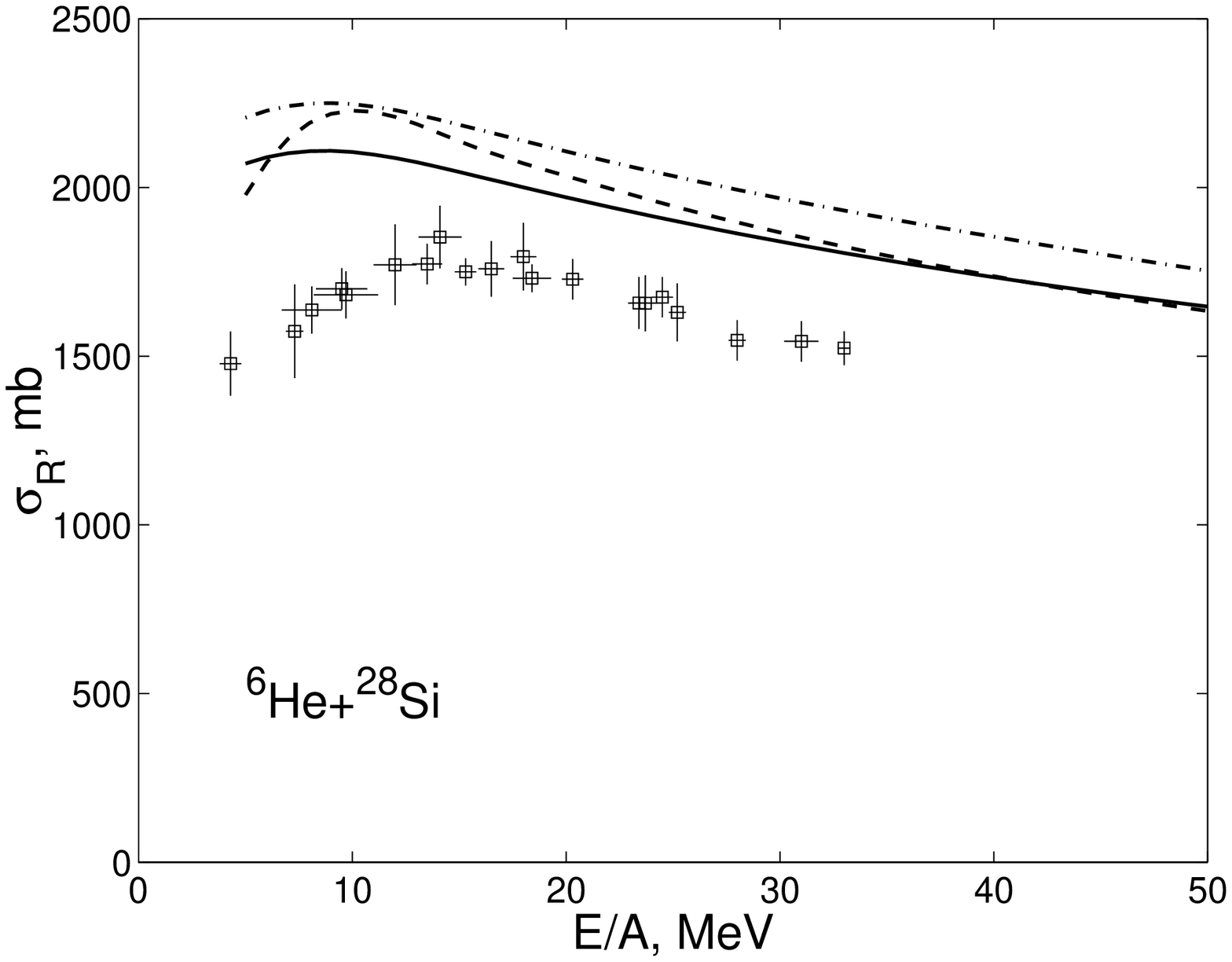, width= .49\linewidth,height=5.cm}
\psfig{file=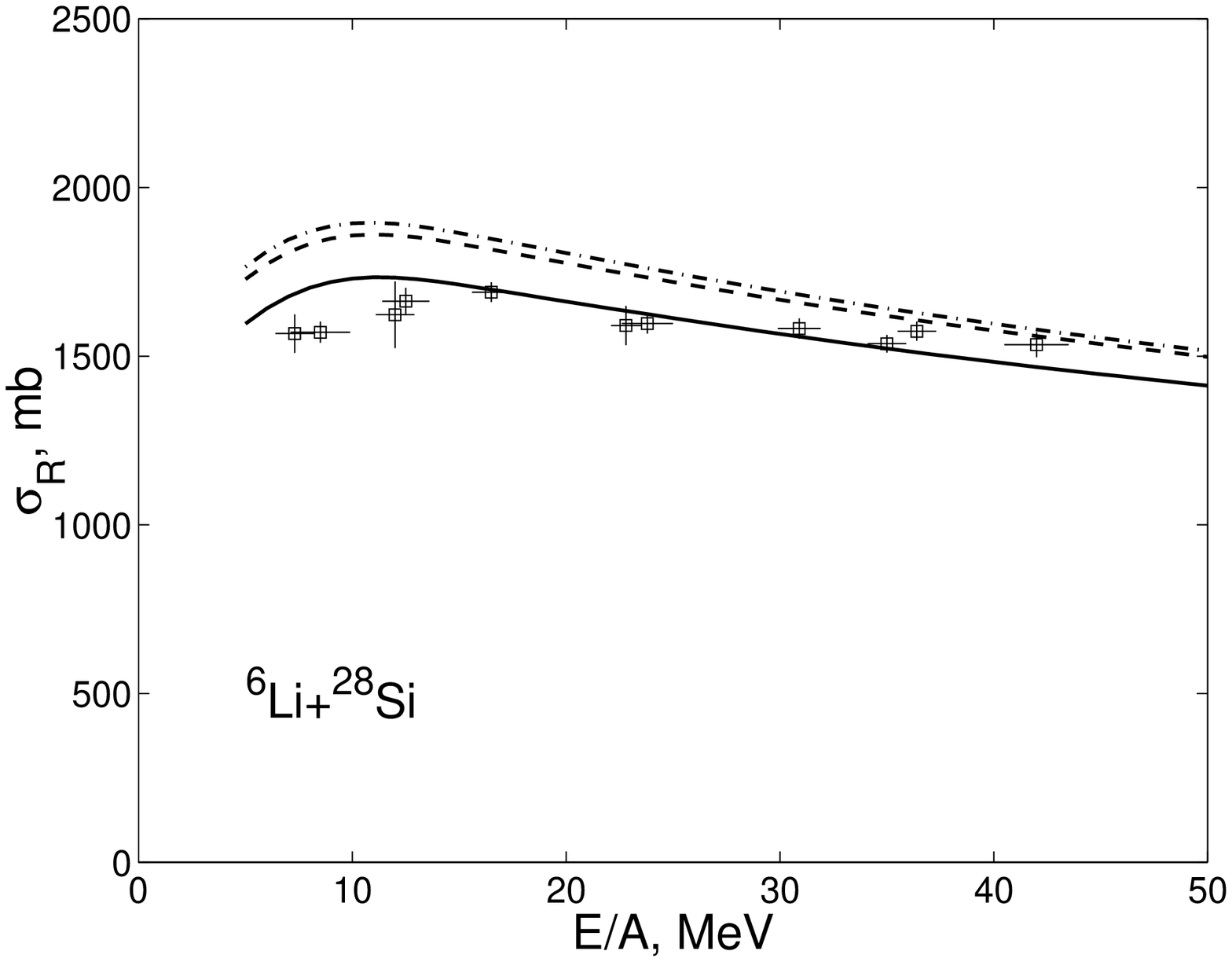, width= .49\linewidth,height=5.cm}
\end{center}
\caption{Microscopic calculations of the total cross sections
using the microscopic optical potential $V^{DF}+iW^H$ without
introducing free parameters. The Paris $NN$-potential is used.
Solid curves: Tanihata density; dashed curves: LSSM; dash-dotted
curves: COSMA.} \label{fig4}
\end{figure}

On the next step of our study, to fit cross sections to the data
we renormalize strengths of the real and imaginary parts of the
potential
%[3.1]
\begin{equation}\label{eq3.1}
U_{opt}(r)\,=\,N_rV^{DF}\,+\,iN_{im}W^H.
\end{equation}
This procedure is commonly used for the real double-folding
potential when one adds the phenomenological imaginary part having
itself several free parameters. Contrastingly, in our study, on
the first stage, we introduce only two parameters to renormalize
strengths of the real $V^{DF}$ and imaginary $W^H$ parts,
calculated microscopically. Fig.~5 shows the results for the
$^6$He+$^{28}$Si cross section when the most realistic LSSM
projectile density of $^6$He was taken in calculations. One sees
that the renormalization makes it possible to agree calculations
to experimental data at larger energies, whereas a significant
discrepancy between the theory and experimental data at lower
energies is still unchanged. We also mention that there exist some
kind of ambiguity when comparing calculated cross sections with
the data. Here we show two nearby curves, the solid one has
renormalization parameters N$_r$=N$_{im}$=0.5, and the dashed one
N$_r$=1.0, N$_{im}$=0.4. So, we conclude that in the framework of
microscopic "volume potentials" (\ref{eq3.1}), the simultaneous
explanation of the data in the whole region of measurements is not
possible. In this connection, at lower energies, the Coulomb
interaction can be thought play a pronounced role in the nuclear
reaction mechanism.

\begin{figure}[t]
\begin{center}
\psfig{file=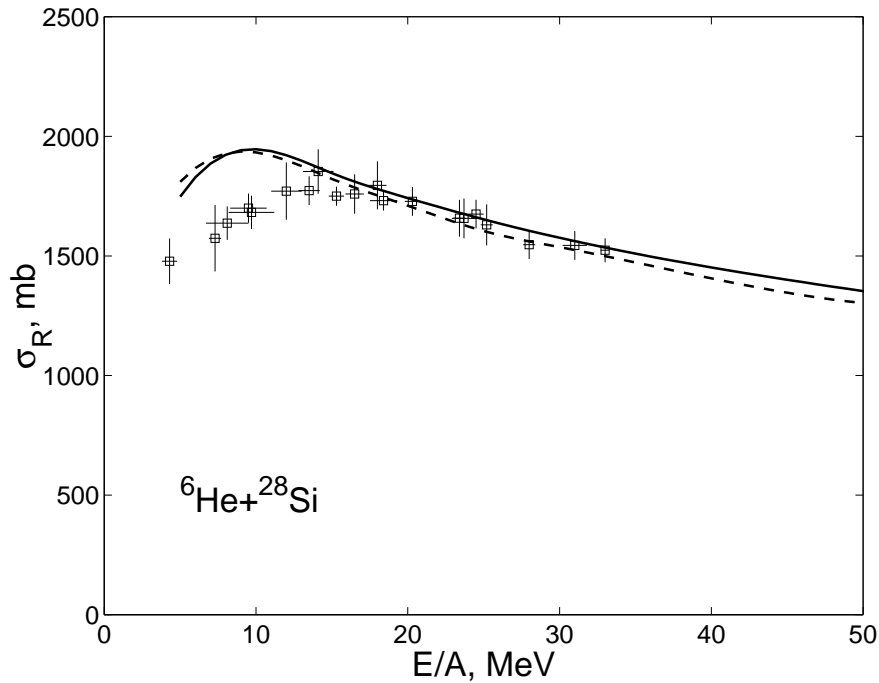, width= .7\linewidth}
\end{center}
\caption{Effect of the strength renormalization of microscopic
potentials $N_rV^{DF}+iN_{im}W^H$ on the total reaction cross
section. Solid curve: $N_r=N_{im}=0.5$; dashed curve: $N_r=1$,
$N_{im}=0.4$.} \label{fig5}
\end{figure}

So, to get more precise results we computed the Coulomb potential
using the microscopic folding formula (\ref{eq2.2}) with the
realistic charge LSSM density and the $NN$ charge interaction
potential $v_C=1/|\bf s|$. Such Coulomb potential and
corresponding cross sections was calculated for the
$^6$He$+^{28}$Si system and compared to that obtained
traditionally with a help of the uniformly distributed charge in
the sphere of the radius of the sum of radii of colliding nuclei.
On the left side of Fig.~6 we exhibit the both potentials. One
sees the visible difference of them in the interior region and
their small separation in the peripheral band, while at larger
distances they coincide to one another. However this changes do
not reveal themselves in behavior of cross sections (right side of
Fig.~6), and the ``bump'' of the $^6$He+$^{28}$Si total reaction
cross section at E$\simeq 15$ MeV is not explained by correcting
the Coulomb potential.

\begin{figure}[t]
\begin{center}
\psfig{file=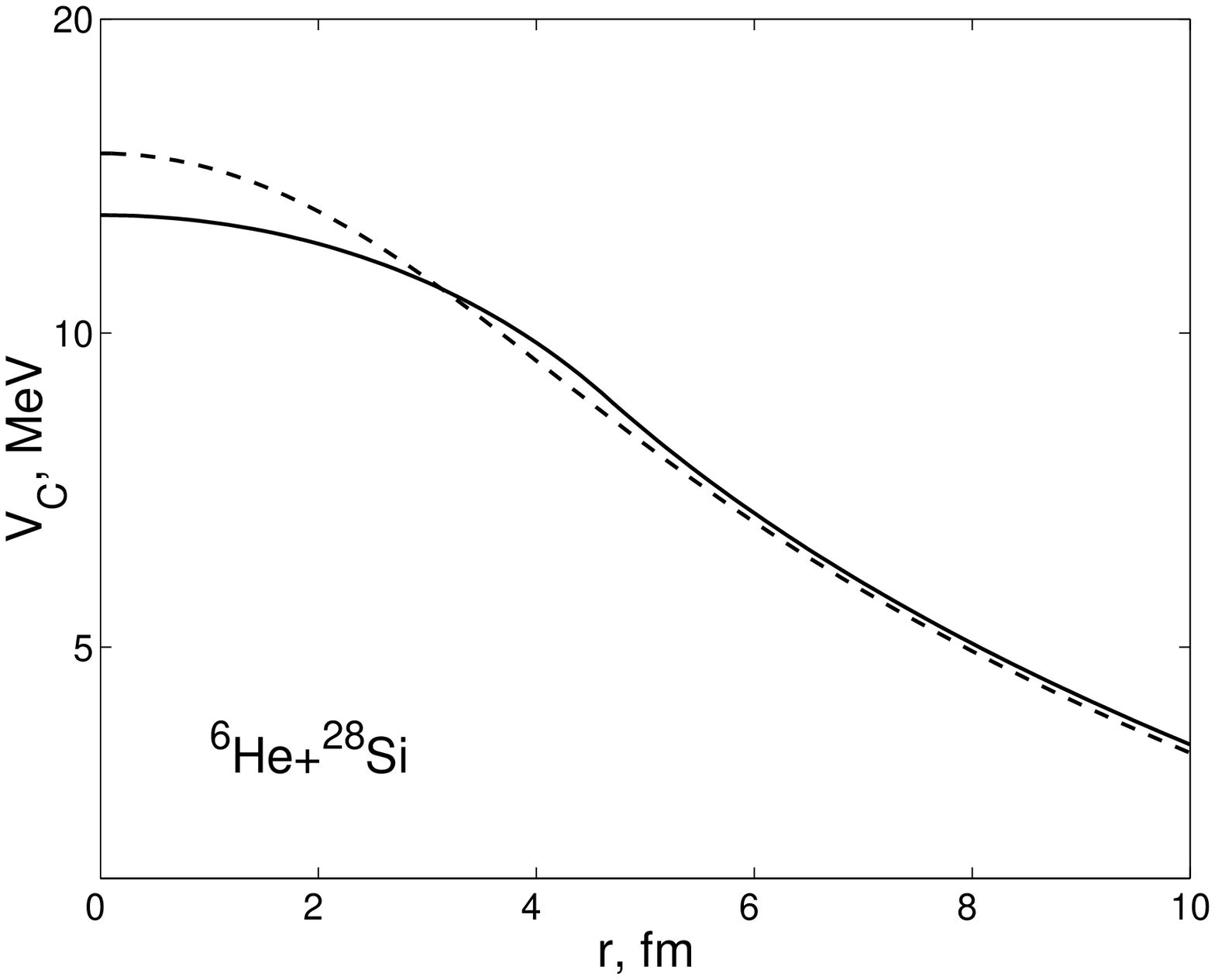, width= .49\linewidth,height=5.cm}
\psfig{file=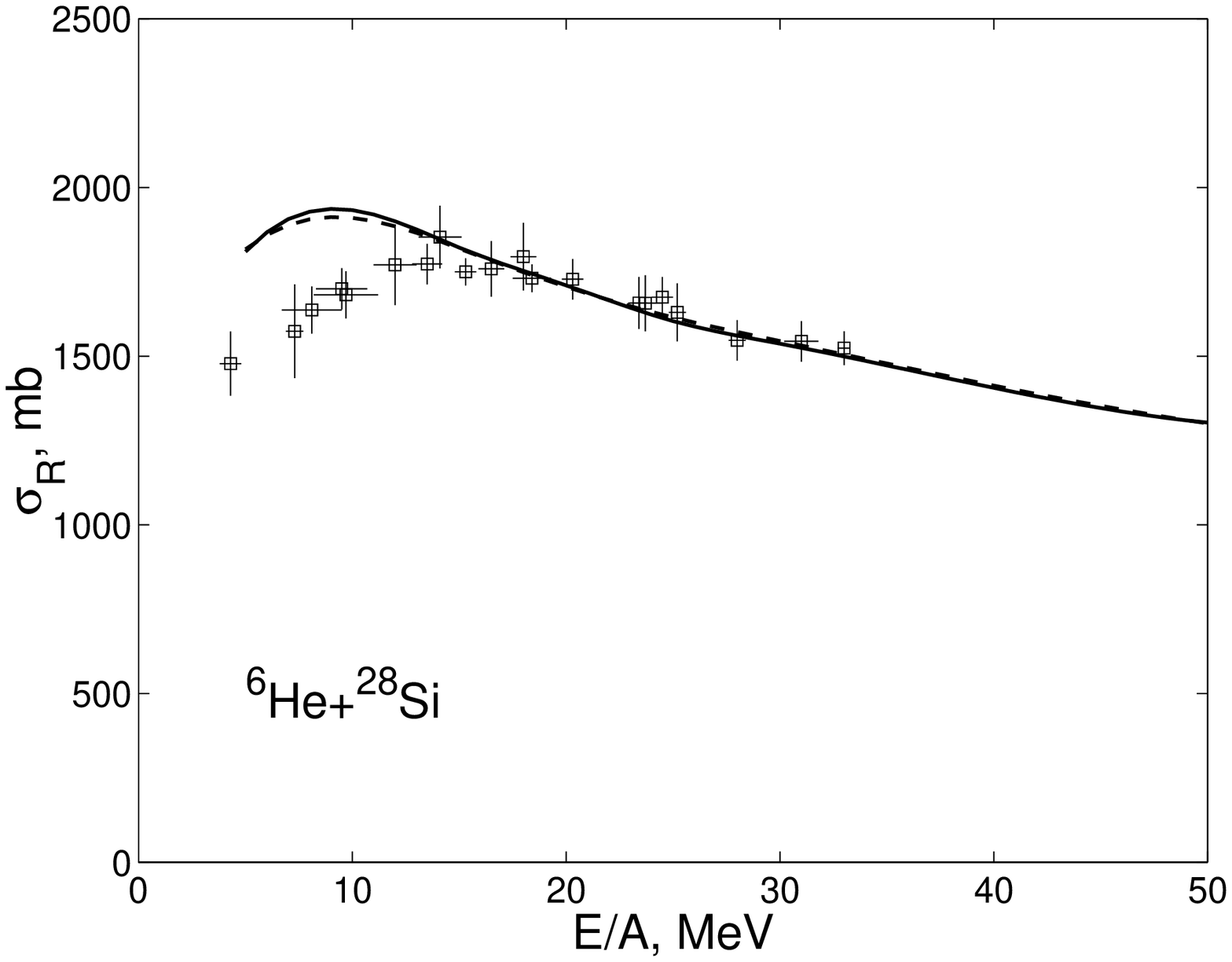, width= .49\linewidth,height=5.cm}
\end{center}
\caption{Effect of correcting the Coulomb AA-potential on the
total reaction cross section. Left panel: the ordinary (solid) and
the corrected (dashed) Coulomb potential for $^6$He$+^{28}$Si.
Right panel: corresponding total cross sections} \label{fig6}
\end{figure}

Going step by step in the framework of our goal to study an applicability of
microscopic potentials, at this stage we intend to simulate an influence
of nuclear collective modes on the mechanism of nucleus-nucleus
scattering. It is known from the theory of inelastic scattering that
excitations of nuclear collective states can be understood by introducing
transition potentials in the form of the derivative of an elastic scattering
potential. With respect to this prescription, we add the derivatives $(-rdV/dr)$
of our microscopic template potentials ("surface terms") to construct
optical potentials
%[3.2]
\begin{equation}\label{eq3.2}
U_{opt}(r)\,=\,\el[N_rV^{DF}-N_r^{(1)}r{dV^{DF}\over dr}\er]\,+\,
i\el[N_{im}W^H-N_{im}^{(1)}r{dW^H\over dr}\er],
\end{equation}
%[3.3]
\begin{equation}\label{eq3.3}
U_{opt}(r)\,=\,\el[N_rV^{DF}-N_r^{(1)}r{dV^{DF}\over dr}\er]\,+\,
i\el[N_{im}V^{DF}-N_{im}^{(1)}r{dV^{DF}\over dr}\er],
\end{equation}
Thus, when fitting cross sections to the data we have two else
free parameters $N_r^{(1)}$ and $N_{im}^{(1)}$ responsible to the
contribution of collective terms. In Fig.7 we demonstrate result
of calculations obtained for two kinds of nucleus-nucleus
potentials. One of them (left panel) is calculated for the Paris
effective NN-potential CDM3Y6 with the LSSM density of $^6$He, and
the other one (right panel) is for the Reid BDM3Y2 NN-potential
\cite{Khoa} with the FDM-model (functional density method)
\cite{FaKukh} of density of $^6$He. The fitted coefficients are
$N_r$=0.7, $N_r^{(1)}$=0.4, $N_{im}$=0.5, $N_{im}^{(1)}$=0.03
(CDM3Y2 case), and $N_r$=1, $N_r^{(1)}$=0.212, $N_{im}$=0.3,
$N_{im}^{(1)}$=0.038 (BDM3Y2 case). It is seen that by introducing
derivatives one can get the fairly well agreement with the
experimental data in the first case, and the qualitative
description for the second potential. These potentials with the
"surface terms" have more smooth diffuseness layers as compared to
"the volume potentials" (\ref{eq3.1}), and this behavior is in
correspondence with the models of collective motions of a nuclear
surface.

\begin{figure}[t]
\begin{center}
\psfig{file=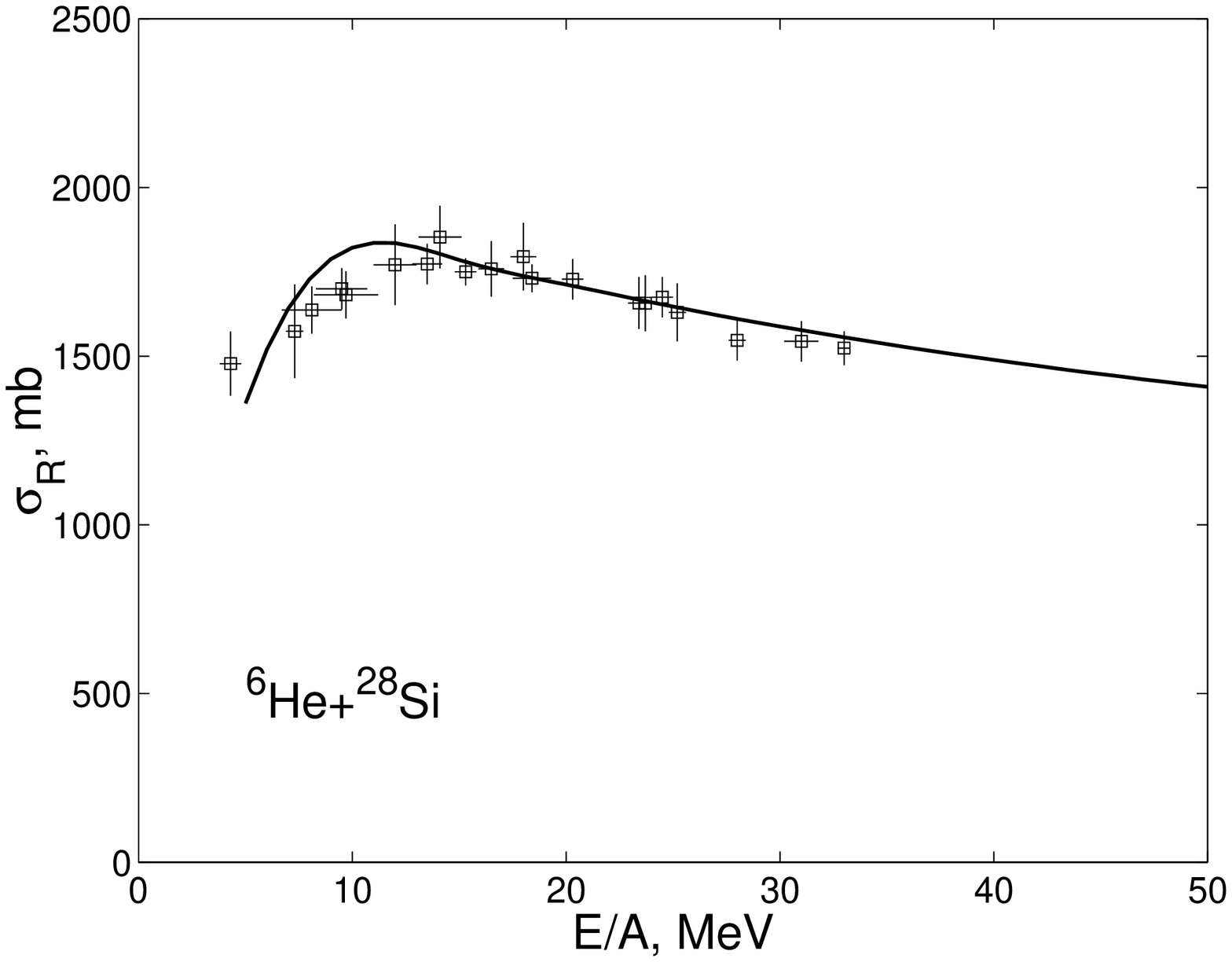, width= .49\linewidth,height=5.cm}
\psfig{file=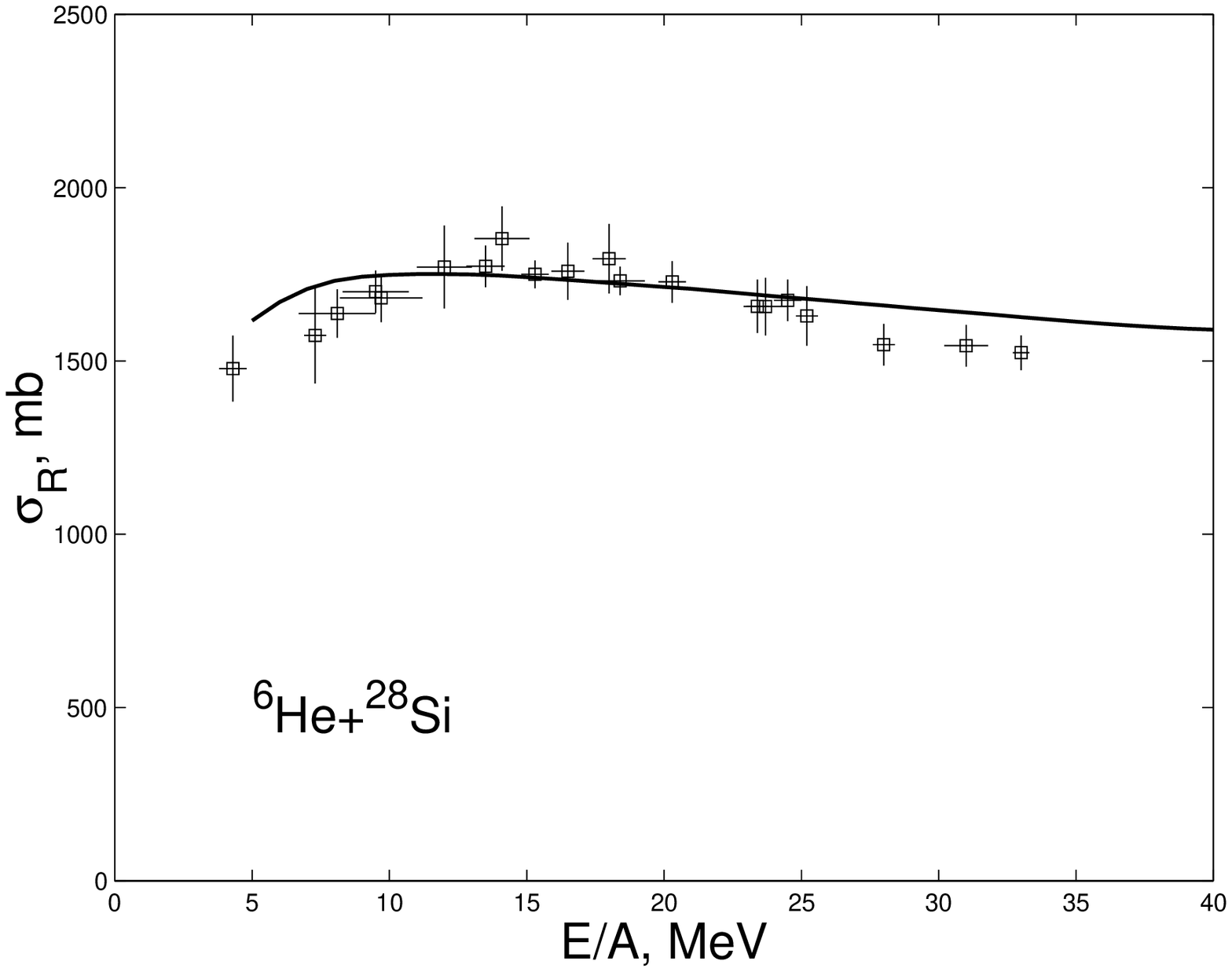, width= .49\linewidth,height=5.cm}
\end{center}
\caption{The fit of the four-parameter semi-microscopic potentials
with the @surface@ terms to the data when used two microscopic
models of the effective NN-forces (see the text).} \label{fig7}
\end{figure}

By the way, such semi-microscopic optical potentials need in further
improvements to exclude free parameters and to give their fully microscopic
interpretation. In this connection, the problem of the physical nature of
an enhancement in the $^6$He+$^{28}$Si total reaction cross section at about
10 MeV over the Coulomb barrier is  still calls for experimental and
theoretical investigations. In particular, the study of angular distributions
in elastic channel can decrease an ambiguity of parameters of semi-microscopic
optical potentials.

\section{Summary and conclusion}

Microscopic models of nucleus-nucleus optical potentials have no
free parameters. They are constructed by using physical
characteristics of  structure of colliding nuclei and  of
effective nucleon-nucleon forces in nuclear medium. We considered
a possibility of the microscopic folding potentials to study the
total cross sections of reactions $^6$He+$^{28}$Si and
$^6$Li+$^{28}$Si. It was shown that a little renormalization of
strengths of this potentials by introducing two parameters allow
for explain the data at comparably higher energies $E \ge 15$
MeV/nucleon. In this region, the cross sections, calculated with
the help of several developed models of the projectile nuclei, are
closely related to each other. Simultaneously, it is seen the
visible disagreement of these calculations with the lower-energy
data, and so that this is the subject of further investigations.
Our treatment to use the microscopically calculated Coulomb
potential does not improve results at these energies. This turn us
to remind that, in general, the ordinary folding potentials take
into account only one-particle density distributions of colliding
nuclei, and thus effects of another channels, connected with
nuclear collective excitations and the nucleon removal reactions,
can also play a role in collisions of nuclei. These effects were
approximately accounted for by adding the derivatives of the
folding potentials to the basic microscopic "volume potential",
and as a result, the fairly well agreement was obtained with the
data at lower energies. Thus one can conclude that the more
developed theory of reactions with exotic beams is called rather
than the use of some kind of phenomenological constructions of
averaged optical potentials based only on methods of
double-folding calculations.

%\section*{Acknowledgments}

The work was partially supported by RFBR (grant 06-01-00228).

\end{document}